\documentclass[openacc]{rstransa}%%%%where rstrans is the template name
\usepackage{subfig}
\usepackage[numbers,sort&compress]{natbib}
\usepackage{float}
\usepackage{fix-cm}
\usepackage{lineno}
\usepackage{aas_macros}
\titlehead{Research}

\begin{document}
%\linenumbers
%%%% Article title to be placed here
%\title{Can Variation in Hydrogen Column Density Be a Signature of Precursors to Coronal Eruptions?}
%\title{Hydrogen Column density: A Probable Signature of Coronal Eruptions}
%\title{Time-Resolved X-ray Spectroscopy of AB Doradus: Tracking $N_H$ Enhancements as Diagnostics for Coronal Eruptions}
\title{Probing Flare-Associated Eruptions on AB Doradus via X-ray Absorption Variations}

\author{%%%% Author details
Shweta Didel$^{1}$,Jeewan C Pandey$^{2}$ and Abhishekh K Srivastava$^{3}$}

%%%%%%%%% Insert author address here
\address{$^{1}$ Central University of Haryana, Mahendragarh - 123031, Haryana, India \\
$^{2}$ Aryabhatta Research Institute of Observational Sciences (ARIES), Manora Peak, Nainital - 263001, Uttarakhand, India \\
$^{3}$ Indian Institute of Technology (Banaras Hindu University), Varanasi - 221005, Uttar Pradesh, India}

%%%% Subject entries to be placed here %%%%
\subject{Astrophysics, Stellar Astrophysics, Solar - Stellar Connection}

%%%% Keyword entries to be placed here %%%%
%%%% Insert corresponding author and its email address}
\corres{Shweta Didel\\
\email{shwetadidel@gmail.com}}

%%%%%%%%%% MY DEF %%%%%%%%%%%%%%%%%%%
\newcommand{\lum}{erg s$^{\rm -1}$\,}
\newcommand{\density}{cm$^{\rm -3}$\,}
\newcommand{\vol}{cm$^{\rm 3}$\,}
\newcommand{\pressure}{dyne cm$^{\rm -2}$\,}
\newcommand{\NH}{{\rm cm$^{-2}$}\,}
\newcommand{\lx}{ L$_{\rm X}$\,}
\newcommand{\lxf}{ L$_{\rm XF}$\,}
\newcommand{\exf}{ E$_{\rm XF}$\,}
\newcommand{\lxq}{ L$_{\rm XQ}$\,}
\newcommand{\nh}{{\rm $N_H$}\,}

%%%% Abstract text to be placed here %%%%%%%%%%%%
\begin{abstract}
In this work, we investigate eruptive phenomena associated with X-ray flares on AB Doradus using observations from XMM-Newton. Our aim is to detect such events with the help of continuous X-ray absorption. The variation in the hydrogen column density (\nh) is used as a primary diagnostic to detect such absorption or spectral hardening in the soft X-ray band below 1 – 2 keV. These variations may represent the signatures of the eruptive processes associated with flares, such as stellar prominences, failed eruptions, or coronal mass ejections. We applied time-resolved spectroscopy technique to track changes in the absorption parameter during flaring episodes. Our analysis shows statistically significant enhancements in the \nh, spanning from 0.3 - 3.4 $\times$ $10^{20}$ \NH during the flare and post-flare intervals. Out of six data sets, two sets containing multiple-overlapped flares show significant fluctuations and remaining showed no such variations in \nh. These variations indicate the presence of dynamic absorbing material along the line of sight and point to ongoing physical processes within the stellar corona. 
\end{abstract}
%%%%%%%%%%%%%%%%%%%%%%%%%%%

%%%%%%%%%% Insert the texts which can accomdate on firstpage in the tag "fmtext" %%%%%

%\begin{fmtext}
%\end{fmtext}
\maketitle
\section{Introduction}
Stellar flares are explosive releases of magnetic energy in stellar coronae, driven by magnetic reconnection in stressed and twisted magnetic field configurations. During a flare, stored magnetic energy is rapidly converted into thermal energy, particle acceleration, and bulk plasma motions, producing enhanced emission from radio to gamma-ray wavelengths \citep{2011LRSP....8....6S,2017LRSP...14....2B}. In magnetically active stars, flares can exceed the most energetic solar events by several orders of magnitude with total energies of $10^{34-36}$ erg. Such events are commonly referred to as superflares \citep{2012Natur.485..478M,2013ApJS..209....5S,2025AJ....169...49D}. These extreme flares are thought to arise from large-scale magnetic structures and complex active regions, analogous to solar two-ribbon flares but on much larger spatial and energetic scales.

On the Sun, flares are frequently associated with eruptive phenomena, including coronal mass ejections (CMEs), prominence or filament eruptions, and large-scale loop expansions. CMEs involve the expulsion of magnetized plasma from the corona into interplanetary space, carrying significant mass and kinetic energy \citep{2011LRSP....8....1C,2012LRSP....9....3W}. Prominences consist of cool, dense plasma suspended in the hot corona by magnetic fields and may erupt partially or fully, sometimes evolving into CMEs \citep{2014LRSP...11....1P}. In contrast, confined or failed eruptions occur when strong overlying magnetic fields inhibit plasma escape, leading to large-scale coronal restructuring without mass loss. Solar observations demonstrate that the probability of CME association increases with flare energy, approaching a near one-to-one correspondence for the most energetic events \citep{Yashiro_2008,2009IAUS..257..233Y,2011SoPh..268..195A}.

Since the first space-based CME observations in the early 1970s \citep{1973spre.conf..713T}, the Sun has remained the only star for which CMEs can be directly imaged. However, extending the study of flare-associated eruptions to other stars is of fundamental importance, particularly in the context of stellar mass loss, angular momentum evolution, and the impact of stellar activity on exoplanetary atmospheres and habitability \citep{2007AsBio...7..167K,2016ApJ...826..195K}. Solar extrapolations suggest that active stars may experience frequent and massive eruptions, and could account for a major share of their total mass loss \citep{2013ApJ...764..170D,2017MNRAS.472..876O,2017ApJ...840..114C}.

Direct detection of stellar CMEs is currently not feasible with existing instrumentation, making it necessary to use indirect observational proxies. Several techniques have been proposed and explored, including Doppler-shifted emission or absorption features in optical and ultraviolet spectra associated with erupting prominences \citep{2019A&A...623A..49V,2022A&A...663A.140L}, radio signatures of large-scale particle acceleration \citep{2018ApJ...862..113C}, transient X-ray absorption by cool or partially ionized plasma \citep{2017IAUS..328..243O,2017ApJ...850..191M}, and coronal dimming phenomena linked to plasma evacuation during eruptions \citep[see][and references therein]{2025LRSP...22....2V}. In the solar context, Mason et al. \cite{2016ApJ...830...20M} demonstrated that the depth and temporal evolution of EUV coronal dimming are closely related to CME mass and propagation speed \cite{2022ApJ...928..154J}. Similar large-scale dimming signatures have also been reported in active stars by Veronig et al. \cite{2021NatAs...5..697V}, who interpreted simultaneous EUV and X-ray dimming as evidence of CME-like plasma eruptions. Subsequent studies by Namekata et al. \cite{2024ApJ...961...23N,2024ApJ...976..255N} further explored coronal eruptions using multiwavelength observations and Doppler-shift diagnostics, providing important constraints on the kinematics, geometry, and eruptive nature of solar and stellar flare-associated plasma motions. These studies suggest that transient absorption, dimming, and complex flare morphologies may provide indirect diagnostics of large-scale coronal restructuring and possible stellar mass ejections. To date, only a limited number of stellar CME candidates have been identified using these methods, highlighting both the rarity of unambiguous detections and the challenges inherent in stellar eruption studies. 

X-ray absorption provides a particularly promising diagnostic for detecting flare-associated mass motions. In the solar corona, erupting filaments and prominences are observed to absorb background emission when cool, dense plasma is lifted into the line of sight \citep{2001ApJ...561..372S,2006ChJAA...6..345J,2012SoPh..277..337V,2016ApJ...821...85G}. In stellar contexts, the significantly larger energies and masses involved in powerful flares may produce detectable changes in the effective \nh. It can be measured over time using time-resolved X-ray spectroscopy. Variations in \nh have been reported during stellar flares and interpreted as signatures of transient absorption by eruptive plasma or large-scale coronal restructuring \citep{2008MNRAS.387.1627P,2012MNRAS.419.1219P,2017ApJ...850..191M,2019ApJ...877..105M}. 

Stellar flares and CME-like eruptive events are important manifestations of magnetic activity that can significantly influence the environments and long-term habitability of exoplanets orbiting active stars \citep{2007AsBio...7..185L}. Energetic phenomena such as flares, stellar winds, and mass ejections collectively contribute to the astrospheric space weather, surrounding the host star, potentially driving atmospheric erosion, enhanced ionization, and climate modification in planetary atmospheres \citep[see][and references therein]{2020IJAsB..19..136A}. Therefore, understanding eruptive activity in young and magnetically active stars such as AB Dor is important not only for stellar astrophysics, but also for astrobiology and studies of star–planet interactions, particularly in the context of terrestrial exoplanets around G, K, and M-type stars.
% we need to understand the nature of global (astrospheric), and local (atmospheric and surface) environments of exoplanets in the habitable zones (HZs) around G-K-M dwarf stars including our young Sun
In this work, we investigate X-ray flares on the active fast rotator K-type star AB Doradus using XMM-Newton observations, focusing on time-dependent variations in \nh as a probe of flare-associated eruptive activity. By comparing the absorption behavior in different flares, we aim to distinguish between confined and eruptive events. The structure of this paper is as follows. In Section \ref{sec:obs}, we describe the observations and the data reduction procedures. Section \ref{sec:results} then details the analysis methods and presents the scientific results derived from the X-ray timing. Finally, in Sections \ref{sec:discussion_nh} and \ref{sec:summary}, we discuss these results and provide our overall conclusions and a summary of the study.

\section{Observation and Data Reduction}
\label{sec:obs}
\begin{figure}
\centering
\vspace{-1.0cm}
\subfloat[S1]{\includegraphics[width=0.45\columnwidth,trim={1.0cm 1.0cm 0.0cm 3.5cm}, clip]{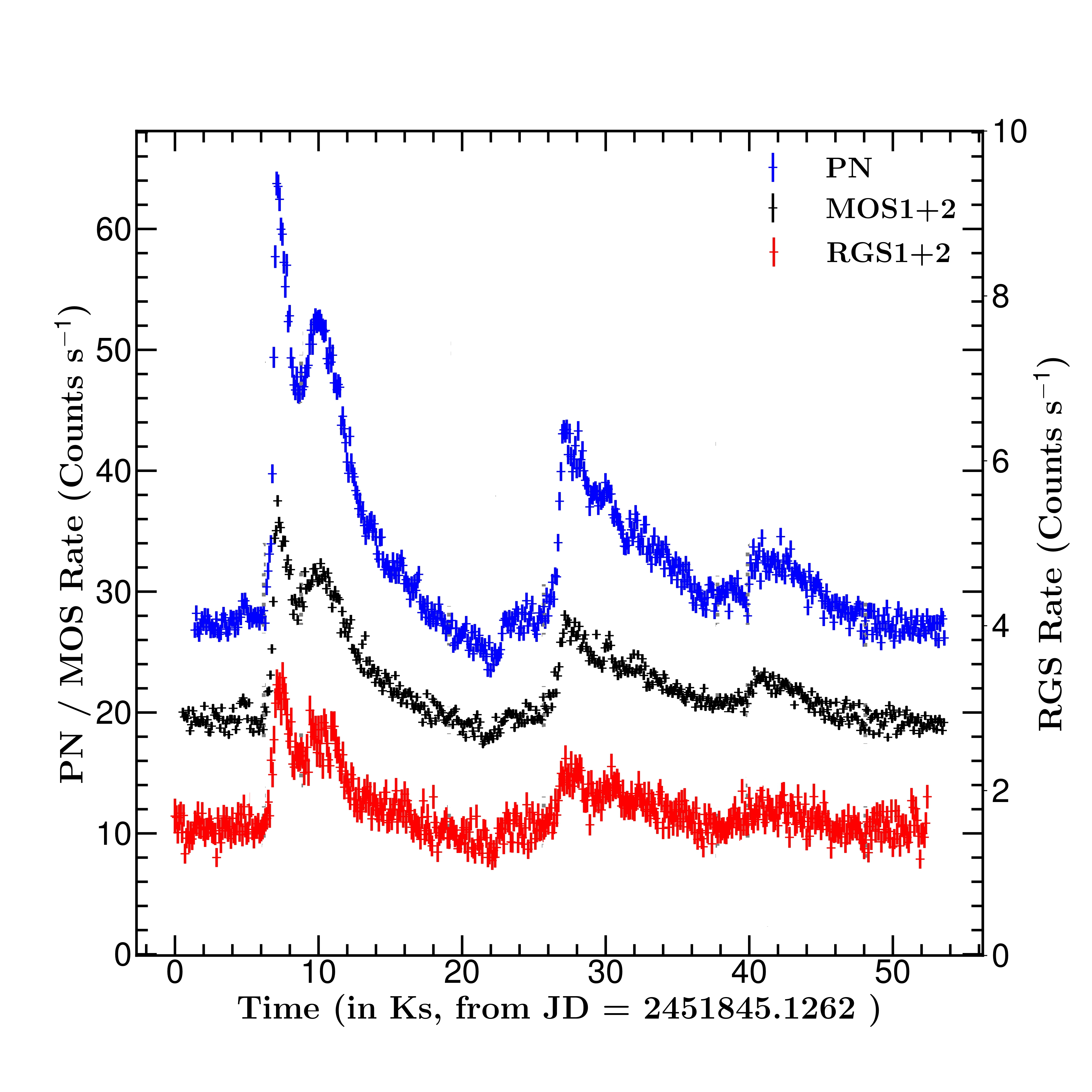}}
\subfloat[S2]{\includegraphics[width=0.45\columnwidth,trim={1.0cm 1.0cm 0.0cm 3.5cm}, clip]{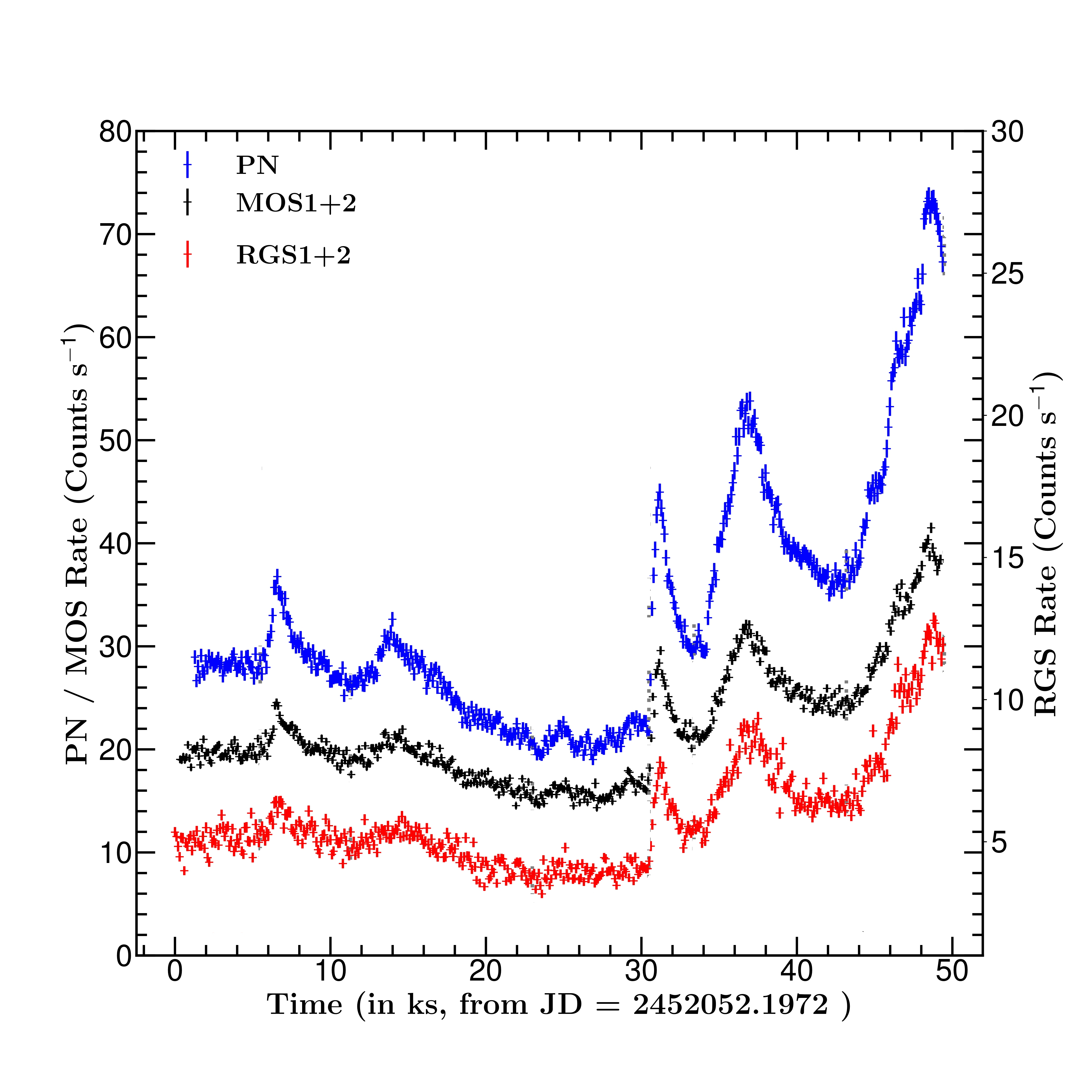}}
\vspace{-0.3cm}
\subfloat[S3]{\includegraphics[width=0.45\columnwidth,trim={1.0cm 1.0cm 0.0cm 3.5cm}, clip]{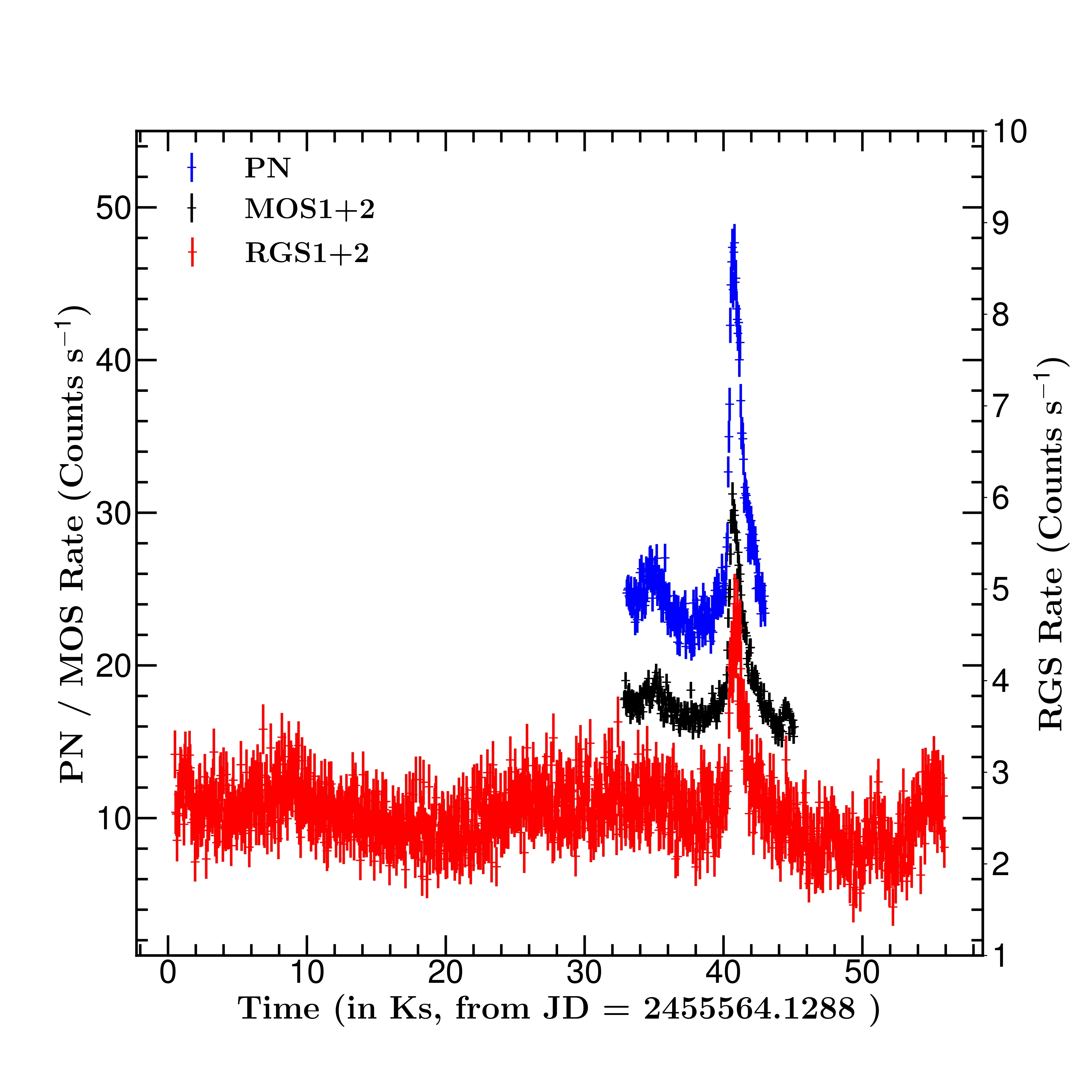}}
\subfloat[S4]{\includegraphics[width=0.45\columnwidth,trim={1.0cm 1.0cm 0.0cm 3.5cm}, clip]{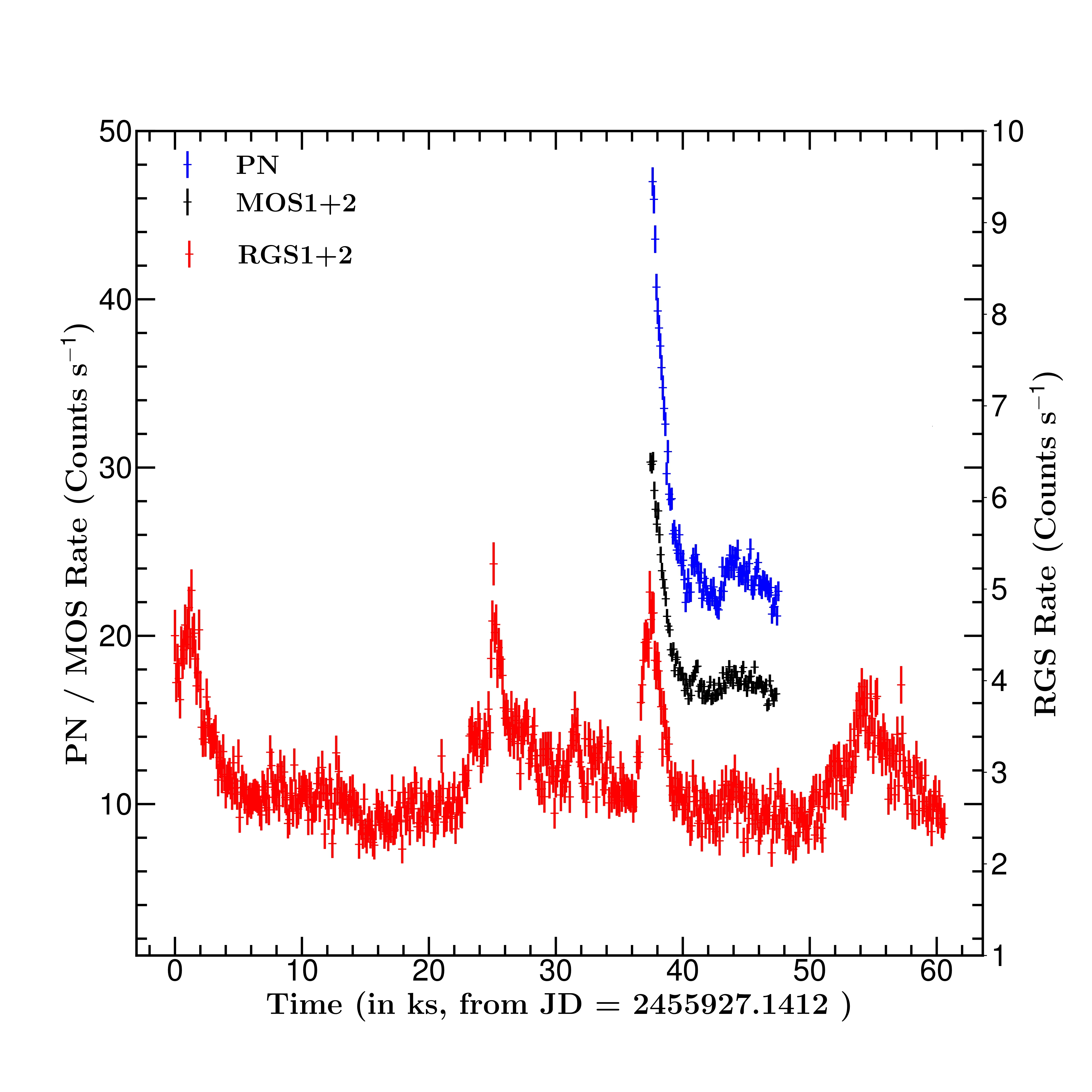}}
\vspace{-0.3cm}
\subfloat[S5]{
\includegraphics[width=0.45\columnwidth,trim={1.0cm 1.0cm 0.0cm 3.5cm}, clip]{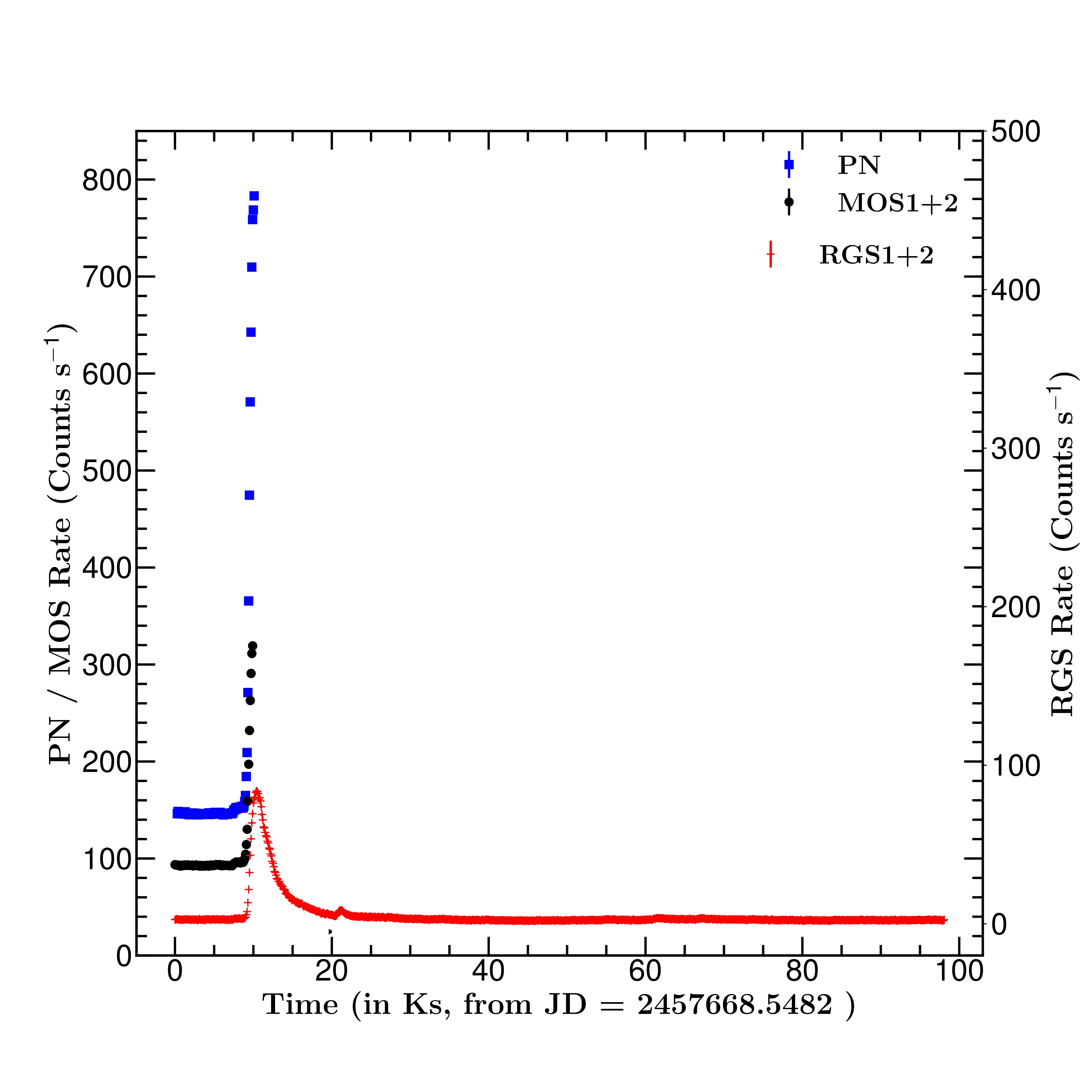}}
\subfloat[S6]{
\includegraphics[width=0.45\columnwidth,trim={1.0cm 1.0cm 0.0cm 3.5cm}, clip]{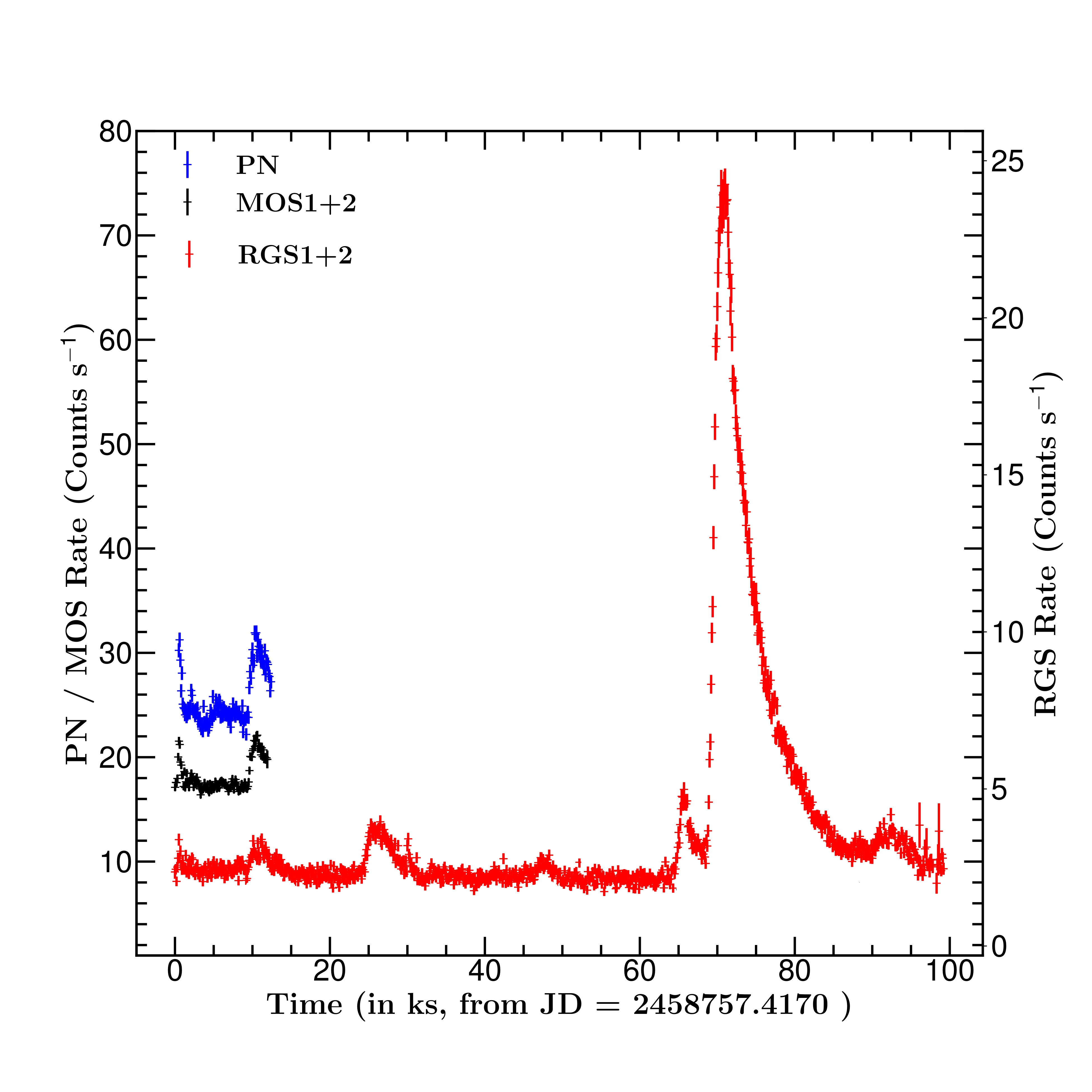}}
\vspace{-0.1cm}
\caption{Background-subtracted X-ray light curves of AB Doradus for different observation epochs. PN, MOS, and RGS data are shown in blue, black, and red, respectively, with count rates scaled for clarity.}
\label{fig:lcs_all_HR}
\end{figure}

This study makes use of the same XMM-Newton observations of AB Doradus analyzed in Didel et al.  \citep{2024MNRAS.527.1705D}. The observation log, including observation IDs, exposure times, and instrumental details, is provided in Table 1 of that work. Data reduction was performed following the identical procedures described in Section 2 of Didel et al. \citep{2024MNRAS.527.1705D}, using standard XMM-Newton Science Analysis System (SAS) tools. All the light curves from total six observations sets are shown in Fig. \ref{fig:lcs_all_HR}.

\section{Analysis}
\label{sec:results}
Our primary aim is to investigate the temporal variation of the hydrogen column density (\nh) across all flaring as well as pre-/post-flare segments. For this purpose, we used time-resolved spectroscopy technique (TRS). We divided each flare light curve into several temporal segments and extracted spectra from each segment. The segments were chosen to have a similar number of counts, which helps to improve the statistical quality of the results. For data sets S1–S4, we used only the PN data because it provides a higher signal-to-noise ratio. For data set S5, both PN and RGS data were used. In contrast, for data set S6 the PN data were insufficient, and the analysis was therefore carried out using RGS data. Each time-resolved PN/RGS spectrum was then modeled using a three-temperature (3T) plasma model \citep{2001ApJ...556L..91S}. Our spectral modeling integrated the {\sc PHABS} (photoelectric absorption) component to characterize \nh, using the solar photospheric abundances of Anders et al. \citep{1989GeCoA..53..197A}, since photospheric abundances of AB Dor are not well established and are commonly assumed to be close to solar values \citep{2018ApJ...862...66W}. In the fitting procedure, \nh was allowed to vary freely along with temperature, emission measure, and elemental abundances of the hottest thermal component, while the temperatures and emission measures of the two cooler components were fixed at their quiescent values, adopted from Table 3 of Didel et al. \citep{2024MNRAS.527.1705D}. In this framework, the first two thermal components represent the quiescent coronal emission, whereas the third component accounts for the flare-related plasma. We adopted this methodology to isolate the true flare contribution by assuming a relatively stable quiescent coronal background. This approach was motivated by the fact that the pre-flare, post-flare, and flare segments consistently showed the two cooler thermal components to remain nearly constant in temperature and emission measure, while the third and hottest component exhibited the dominant variability associated with flare evolution. The average temperatures of the quiescent corona of AB Dor were found to be 3.36 MK and 11.02 MK, with corresponding emission measures of $4.8 \times 10^{52}$ and $6.0 \times 10^{52}$ \density, respectively. To estimate the intrinsic X-ray emission, we additionally included the {\sc cflux} convolution model to derive the unabsorbed flux, which was subsequently used to calculate the X-ray luminosity (\lxf) for each flare segment. All the parameters are shown in Table \ref{tab:all_flares_2+1apec_NH}. The resulting \nh values were found to vary from 0.3 - 3.4 $\times$ $10^{20}$ \NH during flares, and the graphical representation of this temporal variation for all observation sets is shown in Fig. \ref{fig:tvspara_nh}.

\begin{figure}
\centering
%\vspace{-0.1cm}
\subfloat[set S1]{\includegraphics[width=0.95\columnwidth, trim={0.7cm 1.0cm 2.0cm 19.2cm}, clip]{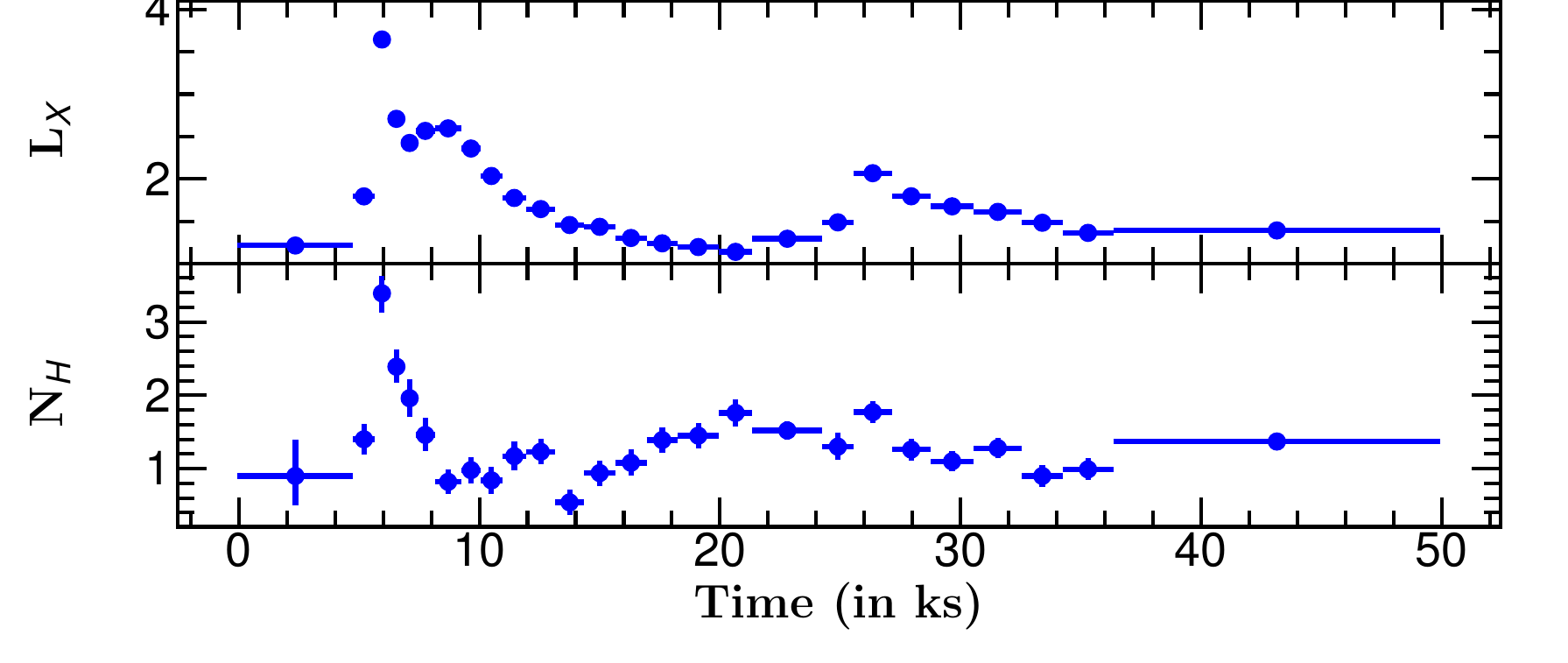}}
\qquad
\subfloat[set S2]{\includegraphics[width=0.95\columnwidth,trim={0.7cm 1.0cm 2.0cm 19.2cm},clip]{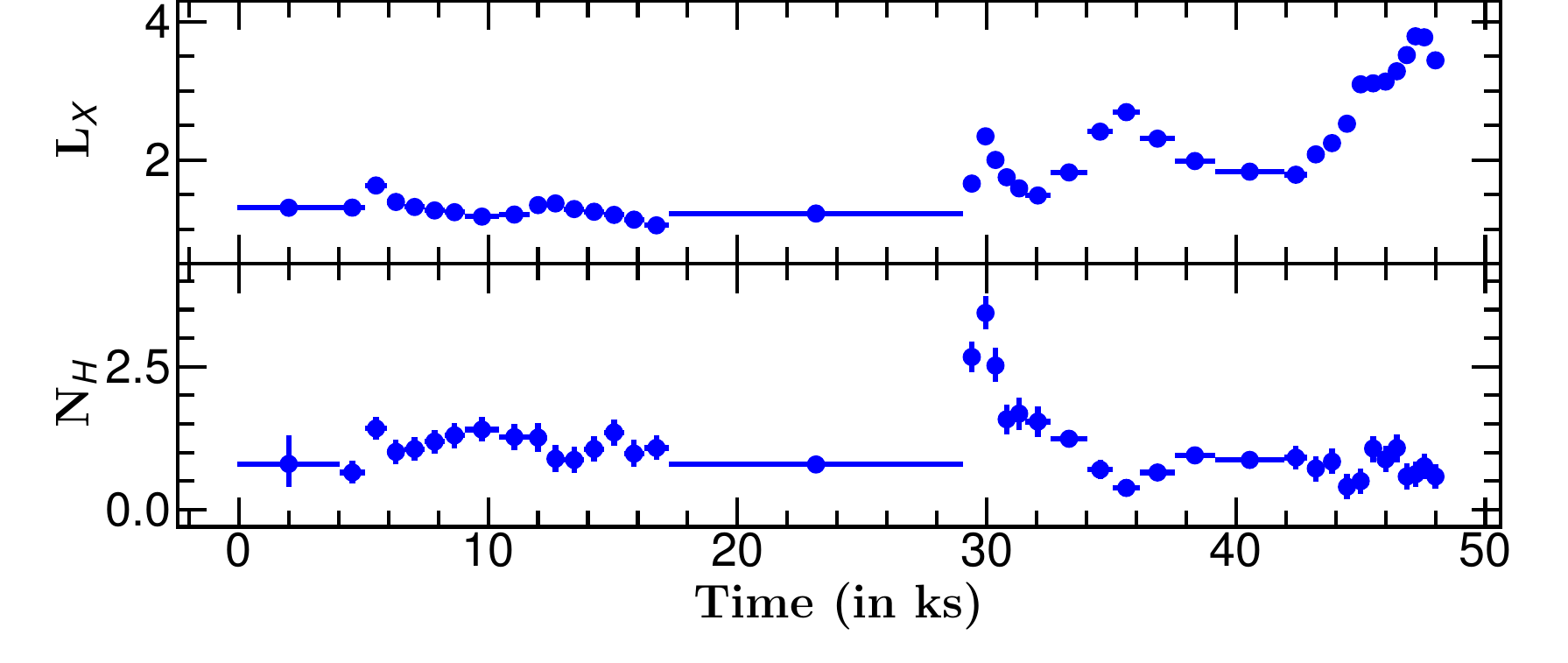}}
\qquad
\subfloat[set S3]{\includegraphics[width=0.95\columnwidth, trim={0.7cm 1.0cm 2.0cm 19.2cm},clip]{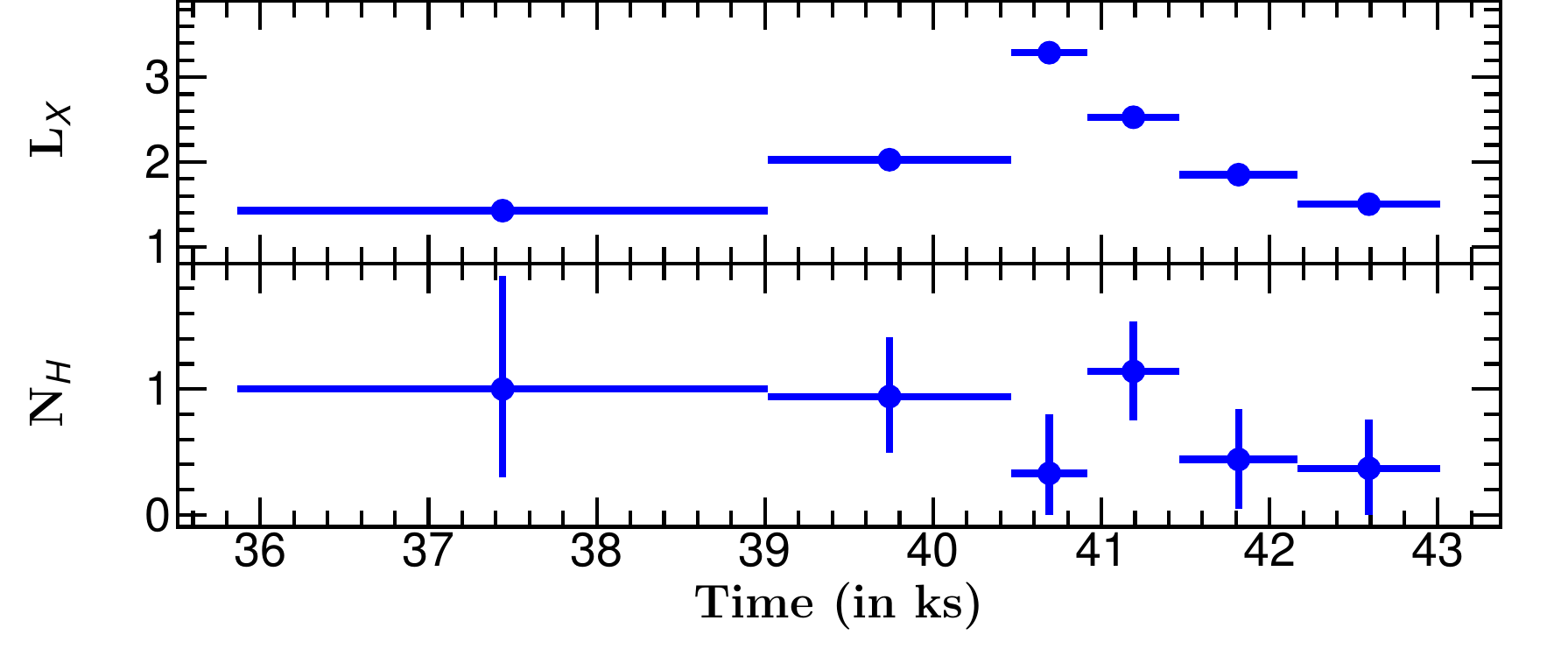}}
\caption{Temporal evolution of the spectral parameters of AB Dor during flares and quiescent states showing X-ray luminosity in units of $10^{30}$ \lum in top panel and hydrogen column density \nh in units of $10^{20}$ cm$^{-2}$ in lower panel. }
\label{fig:tvspara_nh}
\end{figure}
\begin{figure}
 \ContinuedFloat
\subfloat[set S4]{\includegraphics[width=0.95\columnwidth,trim={0.7cm 1.0cm 2.0cm 19.2cm},clip]{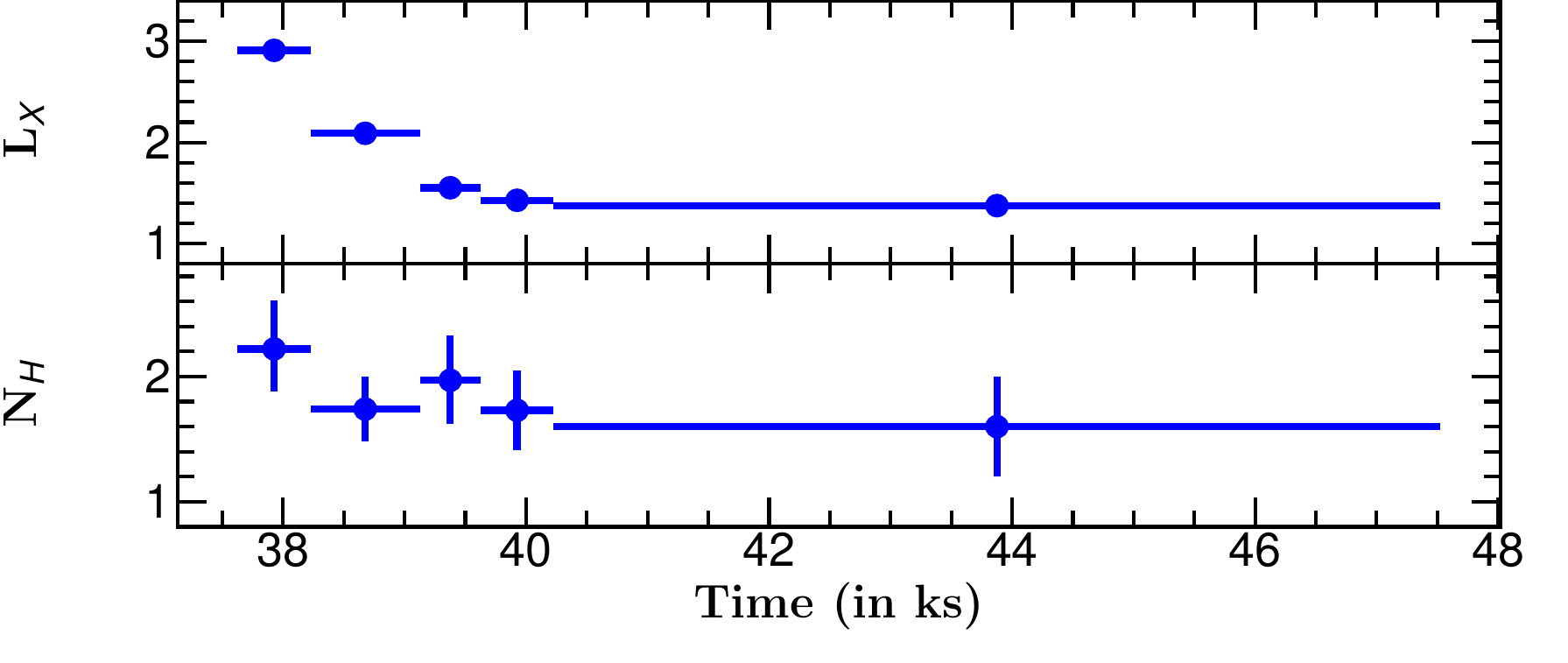}}
\qquad
\subfloat[set S5]{\includegraphics[width=0.95\columnwidth,trim={0.7cm 1.0cm 2.0cm 18.98cm},clip]{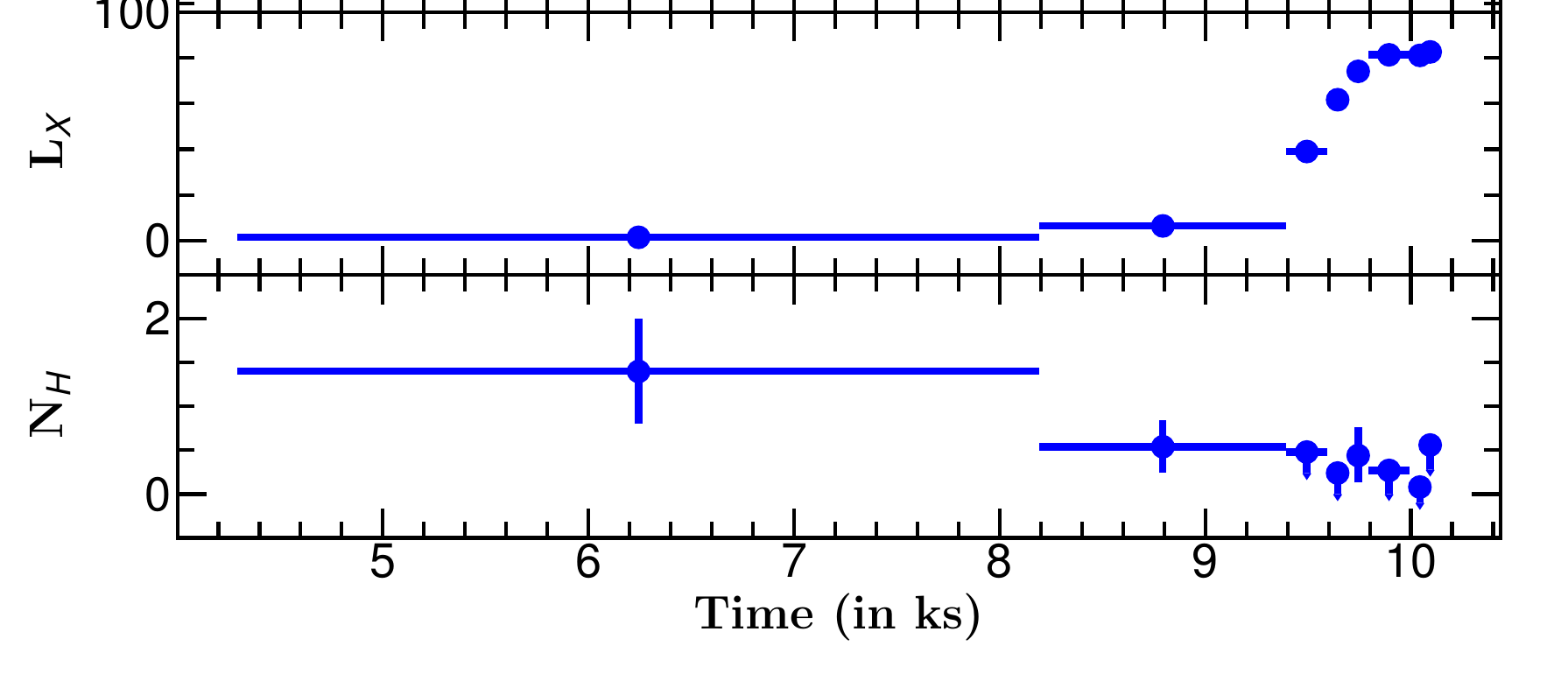}}
\qquad
\subfloat[set S6 ]{\includegraphics[width=0.95\columnwidth,trim={0.7cm 1.0cm 2.0cm 16.65cm},clip]{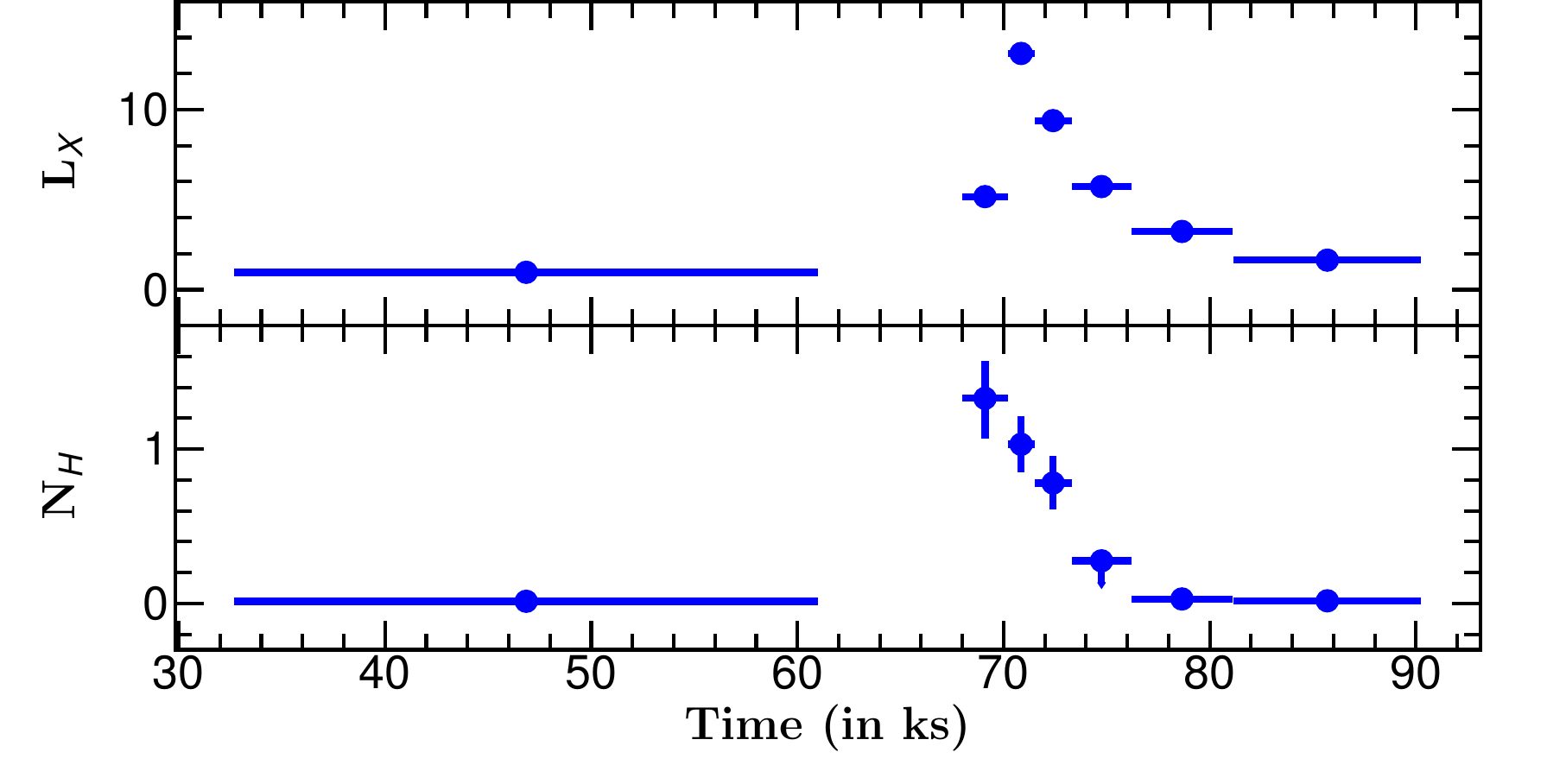}}
%\vspace{-0.25cm}
\caption{Continued.... }
\label{fig:tvspara_nh}
\end{figure}

\section{Results and Discussion}
\label{sec:discussion_nh}
%%%%%1. types of flares ....physics
The analyzed data sets reveal two distinct classes of flares with fundamentally different coronal implications. In Sets S3–S6 of Fig. \ref{fig:lcs_all_HR} [c-f], the flares exhibit a classical rise–peak–decay profile, which is characteristic of energy release within a single dominant magnetic structure, such as an isolated coronal loop or a compact arcade. In these events, the rise phase corresponds to impulsive magnetic reconnection and rapid plasma heating, the peak phase occurs when the loop density reaches its maximum, and the decay phase is governed by radiative and conductive cooling \citep{2024LRSP...21....1K}. In contrast, sets S1–S2 of Fig. \ref{fig:lcs_all_HR} [a-b] display complex, overlapping flares with multiple peaks, where the decay phase of one flare overlaps with the rise of subsequent events. Such behavior is indicative of either sequential or simultaneous flaring in multiple, magnetically connected loops or arcades, reflecting an active region undergoing a sequence of magnetic reconfiguration events or simultaneous flaring at two different locations \citep{2014ApJ...797..122D,2025ApJ...984..186Y}. This prolonged and repeated energy release suggests more towards a highly stressed and dynamic corona in which reconnection in one structure destabilizes neighboring magnetic systems, preventing the plasma from returning to quiescence between events \citep{2014ApJ...797..122D,2014ApJ...797..121H}. Physically, this scenario is analogous to solar flare storms and is expected to enhance the likelihood of large-scale magnetic eruptions, including loop expansion, partial opening, or eruptive mass motions. Consequently, these two flare classes represent intrinsically different coronal regimes: confined, stable loop-dominated flares versus complex, multi-loop events associated with large-scale coronal restructuring. Additionally, using the total energies of the identified flares as mentioned in the Table \ref{tab:all_flares_2+1apec_NH}, we derived the flare energy distribution, $dN/dE \propto E^{-\alpha}$, and obtained a power-law index of $\alpha$ = 1.59 $\pm$ 0.03 \citep{2023RAA....23h5017A}. Although this value is relatively low, it falls within the range reported in previous stellar flare studies \citep[see][and references therin]{2023RAA....23h5017A,2023RAA....23j5010A} and implies that the total energy budget is dominated by the most energetic flares  ($\alpha < 2$). However, we note that the present result is based on a limited flare sample, and a larger statistical dataset is required to constrain the flare energy distribution and assess its significance.
%%%%%%%%Nh variation define and cases possible for such variations

In X-ray spectral analysis, the \nh represents the effective column of cool or partially ionized material along the line of sight and governs the attenuation of soft X-ray photons ($\leq$1-2keV) through photoelectric absorption. Such variability can therefore be interpreted as a signature of dynamic coronal processes involving mass motions, including chromospheric plasma uplift during the impulsive phase, the destabilization and eruption of prominence-like structures, or CME-like events in which dense material is temporarily injected into or crosses the line of sight.

%%%%%%%%%%%pattern of nh variation in each flare and postflare and reason behind it
Time-resolved spectral analysis reveals markedly different absorption behavior between the two flare classes. In Sets S1–S2, which are characterized by complex, overlapping flares, the \nh exhibits significant temporal variability, increasing during the flare rise phase, reaching maximum values near peak luminosity, and subsequently decreasing during the decay as shown in Fig. \ref{fig:tvspara_nh} [a\&b] around 6 ks and $\sim$26 ks in set S1, and near 30 ks in set S2. In several cases, an additional enhancement of \nh is observed during the post-flare phase, particularly around 21 ks in S1 and $\sim$10 ks in S2. To examine whether the observed \nh variations are statistically significant, we further performed a constant \nh $\chi^2$ variability test using the time-resolved spectral measurements and their associated uncertainties. The resulting reduced $\chi^2$ (>> 2) and corresponding p-values (<< 0.001) indicate that the sets S1 and S2 significantly deviate from the hypothesis of constant absorption, supporting the presence of temporal variability in \nh. On the other hand, no statistically significant variability was detected in the remaining flare sets.

Based on the solar CME data, the \nh was found to increase during the flare or post-flare phases and depends on the separation angle between CME and associated flare, location of CME, and line of sight \citep[][]{2008ApJ...673.1174Y,2011SoPh..268..195A}. Such flare-correlated \nh evolution indicates the presence of transient absorbing material associated with the flare itself. Most plausibly arising from flare-driven mass motions such as the uplift of cool or partially ionized plasma during magnetic reconnection, prominence-like material destabilization, or mass eruptions like in CME. As the flare evolves, this material likely expands, becomes progressively ionized, or moves out of the line of sight, leading to a reduction in the effective absorbing column. %It may also be considered as the observed \nh enhancement may be associated with coronal restructuring or partially confined eruptive activity in which the erupting plasma either fails to escape or does not pass through the observer’s line of sight. %coronal restructuring or diagnostic to the eruptive events, where the failed eruption (plasma falls back or remains confined) or ejected plasma doesn't come into line of sight. or The observed \nh enhancement may be associated with coronal restructuring or partially confined eruptive activity in which the erupting plasma either fails to escape or does not pass through the observer’s line of sight.
Post-flare \nh enhancements further suggest delayed mass loading or reconfiguration of overlying magnetic structures, possibly involving cooling of plasma or fallback of erupted material within post-flare arcades. In contrast, Sets S3–S6, which display classical single-peaked flares, show no significant variation in \nh throughout the flare evolution, implying that the absorbing column density along the line of sight remains nearly constant, or that any erupting/ejected material does not significantly intersect the observer’s line of sight. The former case is consistent with confined flares occurring in stable, closed magnetic loops, where plasma heating and chromospheric evaporation remain largely contained and do not produce substantial amounts of cool material intersecting the observer’s line of sight. %The selective presence of \nh variability in S1–S2, combined with its absence in S3–S6, therefore points to fundamentally different coronal regimes: dynamically evolving, multi-loop active regions prone to large-scale restructuring and eruptive or semi-eruptive events versus stable, localized magnetic configurations dominated by confined energy release.

The \nh during all the data sets was found to vary from 0.3 $\times$ $10^{20}$ to 3.4 $\times$ $10^{20}$ \NH. The extinction E(B-V) in the direction of AB Dor is estimated to be 0.07 \citep[][]{2011ApJ...737..103S}. However, Gaia parallax of AB Dor suggests a distance of only 15 pc, so it is probable that the reddening of its light is insignificant or nonexistent, which is also evident from the maximum estimated value of E(B-V) for AB Dor as 0.0003 mag \citep{2018A&A...616A.108B}. The derived value of galactic \nh for AB Dor is also found to be one to two orders of magnitude lower than the \nh derived from X-ray spectral fitting, approximately of the order of $10^{18}$ \NH. This suggests that the value of \nh obtained from X-ray spectra is due to local effects. %However, We note that partial coupling between metallicity (Z), differential emission measure evolution, and soft X-ray absorption (\nh) may influence the spectral fitting, particularly in the low-energy regime where Fe-L complexes contribute significantly to the continuum and line structure. Therefore, some fraction of the observed \nh variability may arise from parameter degeneracies associated with evolving plasma conditions and abundance uncertainties. However, the fact that enhanced \nh is not uniformly observed across all flare sets, despite applying the same fitting methodology throughout the analysis, suggests that the detected absorption variability in the complex flare events is unlikely to be solely an artifact of the fitting procedure.

However, Temporal variations in \nh inferred from time-resolved X-ray spectroscopy can, in principle, arise from spectral fitting artifacts caused by degeneracies among absorption and plasma temperature (kT), particularly in low-count spectra. Because increased \nh and higher kT both suppress soft X-ray photons, poorly constrained fits may artificially enhance \nh to compensate for spectral hardening \citep[e.g.][]{2000ApJ...542..914W}. In addition, partial coupling between metallicity, differential emission measure evolution, and soft X-ray absorption may influence the fitting in the low-energy regime where Fe-L complexes contribute significantly to the continuum and line structure. Therefore, some fraction of the observed \nh variability may arise from parameter degeneracies associated with evolving plasma conditions and abundance uncertainties. %However, the fact that enhanced \nh is not uniformly observed across all flare sets, despite applying the same fitting methodology throughout the analysis, suggests that the detected absorption variability in the complex flare events is unlikely to be solely an artifact of the fitting procedure. 
%Instead, enhanced \nh is preferentially observed only during the morphologically complex flare events, while the remaining flare sets remain broadly consistent with nearly constant absorption within uncertainties despite being analyzed using the same fitting procedure. This suggests that the detected absorption variability is unlikely to arise solely from fitting artifacts, although higher signal-to-noise and multiwavelength observations will be required to more robustly disentangle the relative contributions of abundance evolution, thermal structure, and genuine transient absorption.

If such degeneracies dominated, \nh would be expected to vary randomly or monotonically, closely track temperature evolution, or exhibit similar behavior across all flares analyzed with identical instrumentation and methodology. These signatures are not observed here in our data sets. Instead, significant \nh variations are detected exclusively in the complex, overlapping flares of Sets S1–S2, while \nh remains constant throughout the classical single-peaked flares in Sets S3–S6, despite comparable evolution in kT and EM. Moreover, in S1–S2 the \nh evolution is phase-locked to the flare light curve, increasing during the rise, peaking near maximum luminosity, and decreasing during the decay, with occasional enhancements during post-flare phases—behavior that does not systematically mirror either temperature or emission measure. This selective and coherent behavior strongly disfavors a purely methodological origin and instead indicates a physical mechanism. Similar flare-correlated \nh enhancements have been reported previously and interpreted as transient absorption by cool or partially ionized plasma associated with flare-driven mass motions in stars such as Algol, late type dwarfs, RS Cvn type binaries etc. \citep[see][]{2008MNRAS.387.1627P,2012MNRAS.419.1219P,2017ApJ...850..191M}. The correlation between \nh and flare luminosity further supports this scenario, as complex flares involve larger coronal volumes and sustained magnetic restructuring, increasing the likelihood of prominence eruptions, expanding loops, or CME-like events injecting absorbing material into the line of sight \citep[e.g.][]{2019NatAs...3..742A,2019ApJ...877..105M}. Additionally, alternative explanations for the observed \nh variability include rotational modulation by stable prominence-like structures, absorption by corotating material, and partial occultation or self-absorption within evolving flare loops \citep{1989MNRAS.236...57C,2008ApJ...688..418G,2022ApJ...934...20S,2024ApJ...966...86S}. These processes can produce changes in the line-of-sight absorbing column without requiring a CME.

An empirical relationship between the stellar flare energy in X-rays and its associated CME mass in units of g is estimated as $M_{CME} (g) = 10^{-1.5\pm0.5} E_{G}^{0.59\pm0.02}$, where $E _{G}$ is the X-ray energy in GOES (1 -- 8 \AA) energy band in ergs \citep{2012ApJ...760....9A,2013ApJ...764..170D}. 
The derived X-ray flux is converted into GOES flux using WEBPIMMS for the derived flare temperatures of AB Dor. The estimated values of $M_{CME}$ for AB Dor are found to be in the range $\sim (1-7) \times 10^{18}$ g for most flare events, while the largest flare (F15) yielded a mass of $\sim1.5 \time 10^{19}$ g. \citep[see Table 4 of][]{2024MNRAS.527.1705D} These values of CMEs are 10 to 100 times more than the most massive solar CME \citep{2009IAUS..257..233Y} and similar to previous estimations of stellar CMEs \citep{2022MNRAS.509.3247K, 2022NatAs...6..241N, 1994A&A...285..489G}. Additionally, a tentative trend is apparent from the estimated CME masses, suggesting a threshold of approximately $\sim 2 \times 10^{18}$ g, above which most flares exhibit enhanced \nh values. In contrast, flares with estimated CME masses below this value generally do not show significant absorption variability. Exceptions to this trend are F9 and F15, which possess estimated CME masses above the threshold but do not exhibit measurable \nh enhancements. This may indicate that the eruptive material did not significantly intersect the observer's line of sight, thereby producing little or no detectable absorption signature.

We have also estimated the absorption using the ice-cream cone model \citep{1984ApJ...280..428F} and equation 9 from Moschou et al. \citep{2017ApJ...850..191M}. All the symbols used in this estimation are also taken from Moschou et al. \citep{2017ApJ...850..191M}. The cone height can be estimated by using the equation $d(t) = (S(t)+R_{AB Dor})/(tan(\omega+\phi)+1)$, where $S(t)$ represents the distance traveled by the plasma in time t. The lower and upper limits on S(t) are assumed as $0.2\times R_{*}$ for flare and $15\times R_{*}$ for dynamical length scale. From Table 4 of \cite{2024MNRAS.527.1705D}, we took the average half loop length as $2\times10^{10}$ cm and the ejected mass ($M_{obs}$ or $M_{CME}$) as $3.24\times10^{18}$ g. The estimated values of \nh ranged from 0.65 to $9.95\times 10^{20}$ \NH for the flare scenario and from 0.33 to $2.20\times 10^{20}$ \NH for the dynamical length scale scenario. These values were found to be consistent with the values obtained through the spectral fitting, as shown in Table \ref{tab:all_flares_2+1apec_NH}.

%Future studies combining time-resolved X-ray spectroscopy with high-resolution Doppler-shift measurements, coronal dimming diagnostics, radio observations of shock-related signatures, and coordinated multiwavelength campaigns will be essential for distinguishing between confined magnetic restructuring, failed eruptions, and genuine mass ejections.
%Cautionary paragraph We note that partial coupling between metallicity (Z), differential emission measure evolution, and soft X-ray absorption (\nh) may influence the spectral fitting, particularly in the low-energy regime where Fe-L complexes contribute significantly to the continuum and line structure. Therefore, some fraction of the observed \nh variability may arise from parameter degeneracies associated with evolving plasma conditions and abundance uncertainties. However, the fact that enhanced \nh is not uniformly observed across all flare sets, despite applying the same fitting methodology throughout the analysis, suggests that the detected absorption variability in the complex flare events is unlikely to be solely an artifact of the fitting procedure. Nevertheless, higher signal-to-noise and multiwavelength observations will be required to more robustly disentangle the relative contributions of abundance evolution, thermal structure, and genuine transient absorption.
\section{Summary and Conclusions}
\label{sec:summary}
In this work, we investigated the nature of flare-associated eruptive activity in the corona of the active K-type star AB Doradus through time-resolved X-ray spectroscopy using XMM-Newton observations. By examining the temporal evolution of the hydrogen column density, \nh, during flares with different morphologies, we identified clear differences between classical single-peaked flares and complex, overlapping flares. While the former exhibit stable absorption consistent with apparently magnetically confined flares, the latter show significant and flare-correlated \nh enhancements, with column densities increasing from $10^{19}$ - $10^{20}$ \NH and, in some cases, persisting into post-flare phases. These absorption signatures are best explained by transient line-of-sight obscuration from cool or partially ionized plasma associated with large-scale coronal restructuring, such as prominence lift-off, failed eruptions, or CME-like mass motions. The selective occurrence of \nh variability, together with its phase-locked behavior and independence from temperature and emission measure evolution, argues strongly against spectral fitting artifacts. Although, we admit that non-detections of \nh variability do not necessarily imply the absence of eruptive activity and can be modest eruptions, eruptions occurring outside the observer's line of sight, or absorption signatures below the sensitivity limit of the present data. Our results demonstrate that soft X-ray absorption is a powerful indirect diagnostic of eruptive phenomena in stellar coronae and provides new observational constraints on the coupling between flares and mass ejection in magnetically active stars. 

In conclusion the results suggest that transient \nh enhancements may provide a useful indirect diagnostic of large-scale coronal restructuring and possible eruptive activity in active stellar coronae. Although the observed behavior is qualitatively consistent with transient line-of-sight absorption by cool or partially ionized plasma associated with eruptive mass motions, the present analysis alone cannot uniquely distinguish between CMEs, failed eruptions, prominence destabilization, or other forms of coronal restructuring. Also, the data sample showing \nh variability is statistically small (Only two data sets S1 and S2). Consequently, the methodology should be regarded as promising but still exploratory.

Future progress will require coordinated multiwavelength observations capable of directly probing plasma dynamics and eruption signatures. In particular, high-resolution Doppler-shift measurements can constrain the kinematics of erupting material, coronal dimming studies can trace plasma evacuation from the corona, and radio observations of shock-related Type II bursts can provide independent evidence for large-scale eruptive events \citep[][]{2023ApJ...948....9I,2024ApJ...961...23N,2024ApJ...963...13C,2025Natur.647..603C}.

%\newpage
%\begin{center}

\begin{longtable}{cccccccc}
\caption{Best fit spectral parameters of each temporal segment of the flare. Here, FS denotes the flare segments, while ST and ET indicate the start and end times of each flare segment relative to the start of the corresponding observation. \lxf and \exf represents the X-ray luminosity and energy in 0.3 - 10.0 keV range, respectively. All the flares and flare segments mentioned in the table below are taken from Fig. 1 of Didel et al. \cite{2024MNRAS.527.1705D}. } \label{tab:all_flares_2+1apec_NH}\\
%\begin{tabular}
\hline
Set  &Flare &FS  & ST:ET (ks) &  \nh                  & \lxf      & \exf & $\chi_\nu^2 $ (dof)\\
      &     &    &            & (10$^{20}$ cm$^{-2}$) & ($10^{30}$ \lum) &  ($10^{33}$ erg)  &   \\
 \endfirsthead

\multicolumn{7}{l}%
{{\bfseries \tablename\ \thetable{} -- continued from previous page}} \\
\hline 
Set &Flare&FS  & ST:ET (ks) &  \nh           & \lxf                 &  \exf & $\chi_\nu^2 $ (dof)\\
    &     &    &            & (10$^{20}$ cm$^{-2}$) & ($10^{30}$ \lum) & ($10^{33}$ erg)  &                \\ 
    \hline

\endhead
\hline \multicolumn{7}{r}{{Continued on next page}} \\ 
\endfoot    

\endlastfoot
          
         \hline     
         S1& F1 & R  &   4.8 : 5.7                 & 1.4$_{-0.2}^{+0.2}$ & 1.79$_{-0.01}^{+0.01}$     & 1.61$\pm$0.01& 1.08 (351) \\
           &    & P  &  5.7 : 6.3                  &  3.4$_{-0.3}^{+0.2}$ & 3.64$_{-0.02}^{+0.02}$    & 2.18$\pm$0.01& 1.19 (425)\\
           &    & D1 &  6.3 : 6.9                  &  2.4$_{-0.2}^{+0.2}$ & 2.71$_{-0.02}^{+0.02}$    & 1.63$\pm$0.01& 1.19 (349)\\
           &    & D2 &  6.9 : 7.4                  &  2.0$_{-0.3}^{+0.3}$ & 2.42$_{-0.02}^{+0.02}$    & 1.21$\pm$0.01& 0.95 (316)\\
           & F2 & R  &  7.4 : 8.2                  &  1.5$_{-0.2}^{+0.2}$ & 2.57$_{-0.02}^{+0.02}$    & 2.06$\pm$0.01& 1.02 (359)\\
           &    & P  &  8.2 : 9.3                  &  0.8$_{-0.2}^{+0.2}$ & 2.60$_{-0.01}^{+0.01}$    & 2.86$\pm$0.01& 1.00 (431)\\
           &    & D1 &  9.3 : 10.1                 &  1.0$_{-0.2}^{+0.2}$ & 2.36$_{-0.01}^{+0.01}$    & 1.89$\pm$0.01& 0.96 (373)\\
           &    & D2 &  10.1 : 11.0                &  0.8$_{-0.2}^{+0.2}$ & 2.04$_{-0.01}^{+0.01}$    & 1.84$\pm$0.01& 1.16 (364)\\
           &    & D3 &  11.0 : 12.0                &  1.2$_{-0.2}^{+0.2}$ & 1.78$_{-0.01}^{+0.01}$    & 1.78$\pm$0.01& 1.28 (349)\\
           &    & D4 &  12.0 : 13.2                &  1.2$_{-0.2}^{+0.2}$ & 1.65$_{-0.01}^{+0.01}$    & 1.98$\pm$0.01& 1.04 (358)\\
           &    & D5 &  13.2 : 14.4                &  0.5$_{-0.2}^{+0.2}$ & 1.457$_{-0.009}^{+0.009}$ & 1.75$\pm$0.01& 1.04 (342)\\
           &    & D6 &  14.4 : 15.7                &  0.9$_{-0.2}^{+0.2}$ & 1.436$_{-0.009}^{+0.009}$ & 1.87$\pm$0.01& 1.10 (348)\\
           &    & D7 &  15.7 : 17.0                &  1.1$_{-0.2}^{+0.2}$ & 1.305$_{-0.009}^{+0.008}$ & 1.70$\pm$0.01& 1.03 (329)\\
           &    & D8 &  17.0 : 18.3                &  1.4$_{-0.2}^{+0.2}$ & 1.241$_{-0.008}^{+0.008}$ & 1.61$\pm$0.01& 1.01 (335)\\
           &    & D9 &  18.3 : 20.0                &  1.4$_{-0.2}^{+0.2}$ & 1.197$_{-0.008}^{+0.007}$ & 2.03$\pm$0.02& 1.04 (327)\\
           &    & D10&  20.0 : 21.4                &  1.8$_{-0.2}^{+0.2}$ & 1.141$_{-0.008}^{+0.008}$ & 1.60$\pm$0.01& 1.11 (309)\\
           & F3 & R  &  24.3 : 25.6                &  1.3$_{-0.2}^{+0.2}$ & 1.489$_{-0.009}^{+0.009}$ & 1.94$\pm$0.01& 1.11 (357)\\
           &    & P  &  25.6 : 27.2                &  1.8$_{-0.1}^{+0.1}$ & 2.07$_{-0.01}^{+0.01}$    & 3.31$\pm$0.02& 1.27 (445)\\
           &    & D1 &  27.2 : 28.8                &  1.3$_{-0.1}^{+0.1}$ & 1.796$_{-0.009}^{+0.009}$ & 2.87$\pm$0.02& 1.24 (413)\\
           &    & D2 &  28.8 : 30.6                &  1.1$_{-0.1}^{+0.1}$ & 1.677$_{-0.008}^{+0.008}$ & 3.02$\pm$0.02& 1.13 (417)\\
           &    & D3 &  30.6 : 32.6                &  1.3$_{-0.1}^{+0.1}$ & 1.612$_{-0.008}^{+0.008}$ & 3.22$\pm$0.02& 1.16 (427)\\
           &    & D4 &  32.6 : 34.3                &  0.9$_{-0.2}^{+0.2}$ & 1.484$_{-0.008}^{+0.008}$ & 2.52$\pm$0.02& 1.04 (385)\\
           &    & D5 &  34.3 : 36.4                &  1.0$_{-0.1}^{+0.1}$ & 1.364$_{-0.007}^{+0.007}$ & 2.86$\pm$0.02& 1.04 (390)\\
         S2& F4 & R  &    4.1 : 5.1                &  0.6$_{-0.2}^{+0.2}$  & 1.311$_{-0.01}^{+0.009}$ & 1.31$\pm$0.01& 1.17 (298)\\
           &    & P  &     5.1 : 6.0               &  1.4$_{-0.2}^{+0.2}$ & 1.63$_{-0.01}^{+0.01}$    & 1.47$\pm$0.01& 1.15 (333)\\
           &    & D1 &     6.0 : 6.7               &  1.0$_{-0.2}^{+0.2}$ & 1.39$_{-0.01}^{+0.01}$    & 0.97$\pm$0.01& 1.09 (280)\\
           &    & D2 &     6.7 : 7.5               &  1.1$_{-0.2}^{+0.2}$ & 1.32$_{-0.01}^{+0.01}$    & 1.06$\pm$0.01& 1.01 (286)\\
           &    & D3 &     7.5 : 8.3               &  1.2$_{-0.2}^{+0.2}$ & 1.27$_{-0.01}^{+0.01}$    & 1.02$\pm$0.01& 1.09 (280)\\
           &    & D4 &     8.3 : 9.1               &  1.3$_{-0.2}^{+0.2}$ & 1.24$_{-0.01}^{+0.01}$    & 0.99$\pm$0.01& 1.28 (278)\\
           &    & D5 &     9.1 : 10.5              &  1.4$_{-0.2}^{+0.2}$ & 1.182$_{-0.009}^{+0.009}$ & 1.65$\pm$0.01& 1.14 (275)\\
           & F5 & R1 &     10.5 : 11.7             &  1.3$_{-0.2}^{+0.2}$ & 1.21$_{-0.01}^{+0.01}$    & 1.45$\pm$0.01& 0.99 (261)\\
           &    & R2 &     11.7 : 12.4             &  1.3$_{-0.2}^{+0.2}$ & 1.35$_{-0.01}^{+0.01}$    & 0.95$\pm$0.01& 1.07 (264)\\
           &    & P  &     12.4 : 13.1             &  0.9$_{-0.2}^{+0.2}$ & 1.37$_{-0.01}^{+0.01}$    & 0.96$\pm$0.01& 1.05 (263)\\
           &    & D1 &     13.1 : 13.9             &  0.9$_{-0.2}^{+0.2}$ & 1.29$_{-0.01}^{+0.01}$    & 1.03$\pm$0.01& 1.05 (274)\\
           &    & D2 &     13.9 : 14.7             &  1.1$_{-0.2}^{+0.2}$ & 1.25$_{-0.01}^{+0.01}$    & 1.00$\pm$0.01& 1.08 (266)\\
           &    & D3 &     14.7 : 15.5             &  1.3$_{-0.2}^{+0.2}$ & 1.21$_{-0.01}^{+0.01}$    & 0.97$\pm$0.01& 1.03 (258)\\
           &    & D4 &     15.5 : 16.3             &  1.0$_{-0.2}^{+0.2}$ & 1.14$_{-0.01}^{+0.01}$    & 0.91$\pm$0.01& 1.18 (250)\\
           &    & D5 &     16.3 : 17.3             &  1.1$_{-0.2}^{+0.2}$ &  1.055$_{-0.008}^{+0.008}$& 1.06$\pm$0.01& 1.17 (258)\\
           & F6 & R  &     29.1 : 29.8             &  2.7$_{-0.3}^{+0.3}$ & 1.66$_{-0.01}^{+0.01}$    & 1.16$\pm$0.01& 1.29 (301)\\
           &    & P  &     29.8 : 30.2             &  3.4$_{-0.3}^{+0.3}$ & 2.34$_{-0.02}^{+0.02}$    & 0.94$\pm$0.01& 1.13 (310)\\ 
           &    & D1 &     30.2 : 30.6             &  2.5$_{-0.3}^{+0.3}$ & 2.00$_{-0.02}^{+0.02}$    & 0.80$\pm$0.01& 1.17 (287)\\
           &    & D2 &     30.6 : 31.1             &  1.6$_{-0.3}^{+0.3}$ & 1.75$_{-0.01}^{+0.01}$    & 0.88$\pm$0.01& 1.00 (289)\\
           &    & D3 &     31.1 : 31.6             &  1.7$_{-0.3}^{+0.3}$ & 1.59$_{-0.01}^{+0.01}$    & 0.80$\pm$0.01& 1.01 (277)\\
           &    & D4 &     31.6 : 32.6             & 1.5$_{-0.3}^{+0.3}$ & 1.49$_{-0.01}^{+0.01}$     & 1.49$\pm$0.01& 1.05 (268) \\
           & F7 & R1 &     32.6 : 34.1             &  1.2$_{-0.2}^{+0.2}$ & 1.8$_{-0.01}^{+0.01}$     & 2.70$\pm$0.02& 1.14 (434)\\
           &    & R2 &     34.1 : 35.1             &  0.7$_{-0.2}^{+0.2}$ & 2.41$_{-0.01}^{+0.01}$    & 2.41$\pm$0.01& 1.05 (453)\\
           &    & P  &     35.1 : 36.2             &  0.4$_{-0.1}^{+0.2}$ & 2.70$_{-0.01}^{+0.01}$    & 2.97$\pm$0.01& 1.19 (488)\\
           &    & D1 &     36.2 : 37.6             &  0.6$_{-0.1}^{+0.1}$ & 2.31$_{-0.01}^{+0.01}$    & 3.23$\pm$0.01& 1.27 (482)\\
           &    & D2 &     37.6 : 39.2             &  0.9$_{-0.1}^{+0.1}$ & 1.984$_{-0.009}^{+0.009}$ & 3.17$\pm$0.02& 1.04 (460)\\
           &    & D3 &     39.2 : 42.0             &  0.9$_{-0.1}^{+0.1}$ & 1.831$_{-0.009}^{+0.009}$ & 5.13$\pm$0.03& 1.12 (457)\\
           & F8 & R1 &     42.0 : 42.9             &  0.9$_{-0.2}^{+0.2}$ & 1.79$_{-0.01}^{+0.01}$    & 1.61$\pm$0.01& 1.03 (327)\\
           &    & R2 &     42.9 : 43.6             &  0.7$_{-0.2}^{+0.2}$ & 2.08$_{-0.01}^{+0.01}$    & 1.46$\pm$0.01& 1.11 (342)\\
           &    & R3 &     43.6 : 44.2             &  0.8$_{-0.2}^{+0.2}$ & 2.24$_{-0.02}^{+0.02}$    & 1.34$\pm$0.01& 1.09 (332)\\
           &    & R4 &     44.2 : 44.8             &  0.4$_{-0.2}^{+0.2}$ & 2.52$_{-0.02}^{+0.02}$    & 1.51$\pm$0.01& 1.09 (360)\\
           &    & R5 &     44.8 : 45.3             &  0.5$_{-0.2}^{+0.2}$ & 3.09$_{-0.02}^{+0.02}$    & 1.55$\pm$0.01& 1.10 (376)\\
           &    & P  &     45.3 : 45.8             &  1.1$_{-0.2}^{+0.2}$ & 3.11$_{-0.02}^{+0.02}$    & 1.56$\pm$0.01& 1.19 (369)\\
           &    & D  &     45.8 : 46.3             &  0.9$_{-0.2}^{+0.2}$ & 3.13$_{-0.02}^{+0.02}$    & 1.57$\pm$0.01& 0.97 (368)\\
           & F9 & R1 &     46.3 : 46.7             &  1.1$_{-0.3}^{+0.2}$ & 3.28$_{-0.03}^{+0.02}$    & 1.31$\pm$0.01& 1.03 (339)\\
           &    & R2 &     46.7 : 47.1             &  0.6$_{-0.2}^{+0.2}$ & 3.52$_{-0.03}^{+0.03}$    & 1.41$\pm$0.01& 1.06 (356)\\
           &    & R3 &     47.1 : 47.4             &  0.6$_{-0.2}^{+0.2}$ & 3.79$_{-0.03}^{+0.03}$    & 1.14$\pm$0.01& 1.27 (371)\\
           &    & P  &     47.4 : 47.8             &  0.8$_{-0.2}^{+0.2}$ & 3.77$_{-0.03}^{+0.03}$    & 1.51$\pm$0.01& 0.92 (366)\\
           &    & D  &     47.8 : 48.2             &  0.6$_{-0.2}^{+0.2}$ & 3.44$_{-0.02}^{+0.03}$    & 1.38$\pm$0.01& 1.02 (355)\\
         S3& F10& R  &   39.04 : 40.49             &  0.9$_{-0.4}^{+0.5}$   & 2.03$_{-0.03}^{+0.03}$  & 2.94$\pm$0.03& 0.89 (208)\\
           &    & P  &   40.49 : 40.94             &  0.3$_{-0.3}^{+0.5}$ & 3.29$_{-0.04}^{+0.04}$    & 1.48$\pm$0.02& 1.04 (230)\\
           &    & D1 &   40.94 : 41.49             &  1.1$_{-0.4}^{+0.4}$ & 2.53$_{-0.03}^{+0.03}$    & 1.39$\pm$0.02& 1.16 (252)\\
           &    & D2 &   41.49 : 42.19             &  0.4$_{-0.3}^{+0.4}$ & 1.85$_{-0.02}^{+0.02}$    & 1.29$\pm$0.01& 1.09 (228)\\
           &    & D3 &   42.19 : 43.04             &  0.4$_{-0.3}^{+0.4}$ & 1.50$_{-0.02}^{+0.02}$    & 1.28$\pm$0.02& 0.94 (216)\\
         S4& F13& D1 &     37.63 : 38.23           &  2.2$_{-0.3}^{+0.3}$   & 2.91$_{-0.02}^{+0.02}$  & 1.75$\pm$0.01& 0.88 (268)\\
           &    & D2 &     38.23 : 39.19           &  1.7$_{-0.3}^{+0.3}$ & 2.09$_{-0.01}^{+0.01}$    & 2.01$\pm$0.02& 1.03 (270)\\
           &    & D3 &     39.19 : 39.63           &  2.0$_{-0.4}^{+0.4}$ & 1.55$_{-0.02}^{+0.02}$    & 0.68$\pm$0.01& 1.16 (239)\\
           &    & D4 &     39.63 :40.23            &  1.7$_{-0.3}^{+0.3}$ & 1.43$_{-0.01}^{+0.01}$    & 0.86$\pm$0.01& 1.08 (243)\\
         S5& F15& R1 &   8.19 : 9.39               &  0.5$_{-0.3}^{+0.3}$   & 6.48$_{-0.05}^{+0.05}$  & 7.78$\pm$0.06& 1.03 (411)\\
           &    & R2 &   9.39 : 9.59               &  <0.5                & 39.0$_{-0.3}^{+0.3}$      & 7.80$\pm$0.06& 1.09 (561)\\
           &    & R3 &   9.59 : 9.69               &  <0.2                & 61.7$_{-0.4}^{+0.4}$      & 6.17$\pm$0.04& 1.14 (557)\\
           &    & R4 &   9.69 : 9.79               &  0.4$_{-0.3}^{+0.3}$ & 74.1$_{-0.6}^{+0.6}$      & 7.41$\pm$0.06& 1.20 (472)\\
           &    & R5 &   9.79 : 9.99               &  <0.3                & 81.3$_{-0.6}^{+0.6}$      & 16.3$\pm$0.1 & 1.05 (532)\\
           &    & R6 &   9.99 : 10.09              &  <0.1                & 81.1$_{-0.7}^{+0.7}$      & 8.11$\pm$0.07& 1.11 (432)\\
           &    & P  &    10.09 : 10.19            &  <0.6                & 82.6$_{-0.7}^{+0.7}$      & 8.26$\pm$0.07& 1.14 (451)\\
           &*F15& R  &    8.70 : 10.10             &  2.9$_{-0.2}^{+0.2}$  & 20.3$_{-0.1}^{+0.1}$     & 28.4$\pm$0.1 & 1.10 (1342) \\
           &    & P  &    10.10 : 10.90            &  2.1$_{-0.1}^{+0.1}$ & 46.5$_{-0.2}^{+0.2}$      & 37.2$\pm$0.2 & 1.27 (1778) \\
           &    & D1 &    10.90 : 12.00            &  2.2$_{-0.1}^{+0.1}$ & 33.6$_{-0.2}^{+0.2}$      & 37.0$\pm$0.2 & 1.63 (1610) \\
           &    & D2 &    12.00 : 14.00            &  1.8$_{-0.1}^{+0.1}$ & 16.24$_{-0.08}^{+0.08}$   & 32.5$\pm$0.2 & 1.79 (1744) \\
         S6&*F20& R  &   68.0 : 70.2               &  1.3$_{-0.3}^{+0.2}$   &  5.16$_{-0.04}^{+0.04}$ & 11.4$\pm$0.1 & 1.10 (656)\\
           &    & P  &   70.2 : 71.5               &  1.0$_{-0.2}^{+0.2}$ &  13.11$_{-0.09}^{+0.09}$  & 17.0$\pm$0.1 & 1.26 (905)\\
           &    & D1 &   71.5 : 73.3               &  0.8$_{-0.2}^{+0.2}$ &  9.38$_{-0.06}^{+0.06}$   & 16.9$\pm$0.1 & 1.63 (924)\\
           &    & D2 &    73.3 : 76.2              &  < 0.28              &  5.72$_{-0.04}^{+0.04}$   & 16.6$\pm$0.1 & 1.71 (962)\\
           &    & D3 &     76.2 : 81.1             &  < 0.03              &  3.23$_{-0.02}^{+0.02}$   & 15.8$\pm$0.1 & 1.60 (947)\\
           &    & D4 &      81.1 : 90.3            &  < 0.02              &  1.64$_{-0.01}^{+0.01}$   & 15.1$\pm$0.1 & 1.32 (948)\\
         \hline
%\end{tabular}

\multicolumn{7}{l}
{\footnotesize * Spectral fitting was carried out using RGS spectra.}\\
\end{longtable}

%\end{fmtext}
%%%%%%%%%%%%%%% End of first page %%%%%%%%%%%%%%%%%%%%%

\ack{This work is based on observations obtained with XMM-Newton, the European Space Agency (ESA) science mission with instruments and contributions funded by ESA Member States and by NASA. Data reduction was performed using the XMM-Newton Science Analysis System (SAS), and spectral analysis was carried out with XSPEC. We gratefully acknowledge the XMM-Newton Science Operations Centre for the acquisition, processing, and archival of the data used in this study. S.D. acknowledges the CSIR funding agency for providing the research grant. A.K.S. acknowledges the support of the ISRO project in facilitating this scientific research.}

%%%%%%%%%% Insert bibliography here %%%%%%%%%%%%%%
%\bibliographystyle{plainnat} % use this to have URLs listed in References
\cleardoublepage
\bibliographystyle{RS}
\bibliography{sample}

\end{document}